\documentstyle[aps,preprint]{revtex}
\begin{document}
\draft

\title {Numerical confirmation of universality of transmission 
micro-symmetry relations in a four-probe quantum dot}

\author{Ming Lei and Hong Guo}

\address{ Department of Physics and\\
Centre for the Physics of Materials\\
McGill University\\
Montr\'eal, Qu\'ebec, H3A 2T8 Canada.}

\date{\today}
\maketitle

\begin{abstract}
We study the crossover behavior of the Hall resistance 
between the integer quantum Hall regime and a regime
dominated by the Aharonov-Bohm oscillations, in a system
of 4-probe quantum dot with an artificial impurity confined inside. 
In a previous study [M. Lei, N.J. Zhu and Hong Guo, Phys. Rev. B. 
{\bf 52}, 16784, (1995)], a peculiar set of symmetry relations 
between various scattering probabilities were found in this 
crossover regime.  In this paper we examine the universality
of this set of symmetry relations using different shapes of
the quantum dot and positions of the artificial impurity.
The symmetry holds for these changes and we conclude that
in this transport regime the general behavior of the 
Hall resistance is determined by the competition of the
quantum Hall and Aharonov-Bohm effects, rather than by the
detailed shapes of the structure.

\end{abstract}
\vspace{0.5in}

\baselineskip 16pt
\newpage
\section{Introduction}

Semiconductor nanostructures are unique in offering the possibility of
studying quantum transport in an artificial potential landscape. This is
the regime of {\it ballistic transport}, in which scattering with
impurities can be neglected and one observes the effect of quantum 
fluctuations. Indeed, various experiments conducted on such systems 
observed many interesting phenomena, such as the Aharonov-Bohm (AB) 
effect~\cite{aharonov}, quantum Hall 
effect\footnote{In this paper, quantum Hall effect means integer 
quantum Hall effect.}~\cite{klitzing1,klitzing2}, quantum chaotic 
scattering\cite{marcus}, and bending resistance. The 
transport properties of a system in this regime can be tailored by 
varying the geometry of the conductor, in much the same way as one 
would tailor the transmission properties of an optical 
waveguide~\cite{houten1}. The formal relation between conduction 
and transmission, known as the Landauer 
formula~\cite{imry1,landauer1,buttiker3}, has demonstrated its real power 
in this context. For example, the quantization of the conductance of a 
quantum point contact~\cite{wees1,wharam1}, which is a short
and narrow constriction in a two-dimensional (2D) electron gas, 
can be understood using the Landauer formula as resulting from the 
discreteness of the number of propagating channels in the constriction.  
Since in the ballistic transport regime the current limiting factor 
is provided by the carrier scattering from the structure boundaries,
in general the behavior of various resistances sensitively depends on the
particular shape of the system.

In mesoscopic physics, one of the important effects is the Aharonov-Bohm
effect~\cite{aharonov}, which illustrates that in a field-free
multiply-connected region of space the physical properties of a system
depend on the vector potential \mbox {\boldmath{$A$}}.  It is well
understood that Aharonov-Bohm effect is about quantum interference 
of an one-dimensional electronic system in the presence of a magnetic field.
On the other hand, the quantum Hall effect is associated with 2D
electronic states in strong magnetic fields. In a semi-classical picture 
these states correspond to carrier trajectories grazing the sample's 
boundary.  Quantum mechanically these become the edge states which 
play an important role in understanding the quantized Hall
resistance\cite{halperin}.  In some sample geometries where 
an effective ring topology is obtained, Aharonov-Bohm type oscillations 
can be observed in certain ranges of magnetic field while quantum Hall
resistance is obtained at stronger fields.  A possible example is the case of
a disk with a hole threading through the middle.  Clearly, when the hole is
almost as large as the disk, we have an effectively 1D ring and AB
oscillation is anticipated; when the hole is much smaller than the disk and
with a strong magnetic field, we recover the annular film sample studied by 
Halperin\cite{halperin} where quantized Hall resistance is explained based on
the role of edge states.  While these limiting cases were well understood and
studied, the situation is less clear when the size of the hole or the 
strength of the magnetic field are such that both the AB and quantum 
Hall effects are present. It is the purpose of this work to further study
this ``crossover'' transport regime for the cases where the system is
connected to the outside by quantum wires forming a scattering junction.

We focus on studying the Hall resistance of several 4-probe device 
structures using the B\"uttiker multi-probe resistance 
formula\cite{buttiker1,buttiker2},
\begin{equation}
R_{mn,kl}\ =\ \frac{h}{e^2}(T_{km}T_{ln}-T_{kn}T_{lm})/D\ \ \ ,
\label{eq1}
\end{equation}
where $T_{mn}$ is the transmission coefficient from probe $n$ to probe $m$,
and the factor $D$ is a subdeterminant of rank three of the matrix formed
by the equations for the electric current\cite{buttiker1}.  Particular 
structures of interests are shown in Figures (\ref{fig1}), (\ref{fig2}) 
and (\ref{fig5}).  For these systems the Hall resistance is given by 
$R_{13,24}$ using Eq. (\ref{eq1}). These systems are of the usual Hall-bar
structure with an antidot or artificial impurity confined inside at 
various positions.  Experimentally these structures can be fabricated
by the recently developed multilevel fabrication technique\cite{feng},
and very interesting anomalous transmission behavior has already
been observed in a 2-probe situation\cite{kirczenow}.
Clearly, when the magnetic field is high, {\it e.g.}, $l_s/l_B >> 1$
with $l_s$ the ``constriction length'' (see Fig. (1)) and 
$l_B=\sqrt{\hbar c/eB}$ the ``magnetic length'', quantum Hall regime
is reached. In this case each perfectly transmitting channel or edge state
contributes a factor of $e^2/h$ to the Hall conductance, thus $N$ such
channels give rise to a quantized conductance $G=N\frac{e^2}{h}$.  On the
other hand, if the size of the antidot is so large such that $l_s/l_B << 1$,
the structure is effectively a ring since the conducting electrons do not
probe the corners of the quantum dot very much.  In this case we expect to 
and indeed observe AB oscillations\cite{lei1}. However when the length 
scales are such that $l_s/l_B \sim 1$, we enter the crossover regime 
and the transmission properties become complicated.

In a previous paper\cite{lei1}, we have made an investigation
of this crossover regime using the device structure of Fig. (1).  
There, we have systematically varied the radius of the antidot and 
the strength of the magnetic field $B$.  We have found that for 
a range of these parameters the transport behavior enters a peculiar
regime where a very interesting ``micro-symmetry'' (see below) exists
between various transmission probabilities.  The consequence of this 
symmetry was the observation of an anomalous Hall plateau before the 
quantum Hall regime is reached, namely an Hall plateau
was observed without the establishment of perfectly transmitting edge
states. Furthermore, we found that this set of symmetry
relations could be explained based on the general topological equivalence
of the dominating transmission patterns\cite{buttiker1}. Clearly, a
further and important question to ask is whether or not this set of
peculiar symmetry relations hold for other shapes of the scattering
junction (such as for that of Fig. (\ref{fig2}))), or for systems without an
obvious 4-fold rotational symmetry (such as for those in Fig. (\ref{fig5})).
Since the topological equivalence argument which ``derives'' the
micro-symmetry relations\cite{lei1} is quite general and does not depend 
on system details, we expect an ``universal'' behavior of the 
scattering probability. Namely, the symmetry relations observed in the 
system of Fig. (\ref{fig1}) should also manifest in other systems 
shown in Figs. (\ref{fig2}) and (\ref{fig5}).  Indeed this is what is 
obtained from our numerical study to be presented below.  Thus we may 
conclude that the micro-symmetry relations obtained in
Ref. \onlinecite{lei1} is general, and is an universal behavior of 
this peculiar crossover transport regime.
 
The paper is organized as the following. In the next section, for
completeness and ease of discussion, we briefly summarize the findings of
our earlier work of Ref. \onlinecite{lei1}.  Section III present our
investigation of the universality of the symmetry relations.  Finally a
summary and discussion is presented in the last section.   

\section{The micro-symmetry}

In order to make the presentation clearer, in this section we briefly 
review the micro-symmetry relationships we have found in a previous 
study for the quantum dot system shown in Fig.~(\ref{fig1}).  
The details have been included in Ref. \onlinecite{lei1} and interested 
readers should refer to that article for complete results.

Consider this structure with the quantum dot size $D=2W$ where $W$ is the
width of the probes, taken to be $1650\AA$.  
For an incoming electron energy $kW=9.5$ 
which is just above the third subband energy when magnetic field 
is absent, let's make sure that there are at least two transmitting 
channels through the system for even the largest antidot radius $r_a$ 
which we study. We have found, as mentioned in the Introduction, that 
for small $r_a$ one obtained a quantized Hall resistance when $B$ is 
such that perfectly transmitting edge states are established from probe 
I to II. On the other hand for a very large $r_a$, one observed typical 
AB oscillations.  For an intermediate range of $r_a$ with a given value 
of $B$, the crossover regime is reached and Ref. \onlinecite{lei1} 
has shown the following concerning this crossover regime:

\begin{itemize}

\item[(1).] If we denote $T_{ij,mn}$ as the probability of an electron 
incoming from probe $n$ at channel $j$, and going into probe $m$ 
at channel $i$, then Ref. \onlinecite{lei1} shows that 
the following ``micro-symmetry'' hold approximately 
\[ T_{11,11}\ =\ T_{22,21}\ \,\ \  T_{22,11}\ =\ T_{11,21}\ \ , \]
\[ T_{21,11}\ =\ T_{12,21}\ \,\ \  T_{12,11}\ =\ T_{21,21}\ \ , \]
\[ T_{11,31}\ =\ T_{22,41}\ \,\ \  T_{22,31}\ =\ T_{11,41}\ \ , \]
\begin{equation}
T_{21,31}\ =\ T_{12,41}\ \ ,\ \ T_{12,31}\ =\ T_{21,41}\ \ \ .
\label{eq7}
\end{equation}
These symmetry relations are nontrivial. In Ref. \onlinecite{lei1}
we have ``derived'' these relations using a topological equivalence
argument based on the dominating transmission patterns.

\item[(2).] Because of this micro-symmetry (\ref{eq7}), we easily 
deduce the following relations between transmission coefficients of 
individual channels:
\begin{equation}
T_{11}^{1}=T_{21}^{2}\ ,\ \ T_{21}^{1}=T_{11}^{2}\ ,\ \
T_{31}^{1}=T_{41}^{2}\ ,\ \ T_{41}^{1}=T_{31}^{2}\ \ ,
\label{eq5}
\end{equation}
which indicate that the reflection coefficient of channel 1, $T_{11}^1$,
equals the transmission coefficient to probe II of channel 2, $T_{21}^2$;  
and that the transmission coefficient to probe III of channel 1, $T_{31}^1$, 
equals the transmission coefficient to probe IV of channel 2, $T_{41}^2$.  

\item[(3).] Moreover, for the structure of Fig. (\ref{fig1}) the 
transmission coefficients happen\cite{lei1} to approximately have the 
following values in the crossover regime:
\[ T_{11}\ \equiv\ T_{11}^{1}+T_{11}^{2}\ =\ 1\ ,\]
\[ T_{21}\ \equiv\ T_{21}^{1}+T_{21}^{2}\ =\ 1\ ,\]
\[ T_{31}\ \equiv\ T_{31}^{1}+T_{31}^{2}\ =\ 0\ ,\]
\begin{equation}
   T_{41}\ \equiv\ T_{41}^{1}+T_{41}^{2}\ =\ 0\ .
\label{eq4}
\end{equation}
Obviously we expect that these particular values are system dependent.

\item[(4).] One consequence of the symmetry relations (\ref{eq7}) and
(\ref{eq5}) is, using the B\"uttiker formula, that Hall resistance $R_H$
takes an apparent plateau (quantized) value before perfectly transmitting
edge states are established.  With the particular values of the transmission
coefficients which are given by (\ref{eq4}), this abnormal Hall
plateau has the value of $1\times \frac{h}{e^2}$.

\end{itemize}

\section{Universality of the micro-symmetry}

In this section, we investigate the universality of the micro-symmetry in
the crossover regime which is summarized in the last section.
Our main task is to answer the following question: is the micro-symmetry 
a universal property of the crossover transport regime without or
with only weak dependence on the particular geometric shape of the sample ?
Although the topological equivalence argument\cite{lei1} seems to give a
positive answer to this question, we are not aware of any analytical
technique which can give a rigorous derivation of this micro-symmetry. The 
difficulty comes from the fact that a complicated 2D scattering problem must
be solved in order to obtain the various transmission probabilities.  
Hence we have decided to numerically examine several different
systems and numerically compute the scattering probabilities to check
the validity of the symmetry relations presented in the last section.
We emphasis that this work is a numerical establishment of the generic
behavior of the crossover regime, rather than a rigorous proof.

The quantum scattering problem is solved using the finite-element numerical
scheme of Ref. \onlinecite{lent,guo1}. Essentially we discretize the scattering
region which is the quantum dot into a fine grid of finite elements.  This 
procedure reduces the solution of the Schr\"odinger equation into a
sparse matrix problem.  The quantum propagation in the probes is solved
separately using the method outlined in Ref. \onlinecite{schult}.  
The wavefunctions and their spatial derivatives are matched at the
probe boundaries, and this leads to the transmission probabilities
and coefficients.  We have checked, for all the systems studied below,
that numerical convergence is obtained using around 5000 grid points.
As shown in our earlier work\onlinecite{lei1} and discussed above, 
the crossover transport regime is reached for a range of appropriate values 
of the anti-dot size $r_a$ and the magnetic field $B$.  To save computation 
requirement in this work we have fixed $B$ at a reasonablely large value
and varied the size $r_a$ to reach the crossover regime.  

\subsection{Circular quantum dot}

First, let's consider the situation where the geometry shape of the quantum 
dot is changed to circular, see Fig. (\ref{fig2}). The antidot is still 
fixed at the center of the (circular) quantum dot.  The parameters $W$,
$D$ and $kW$ are the same as those used in Ref. \onlinecite{lei1} and
quoted in the last section. To investigate the crossover regime,
we fix the magnetic field $B$ at $6060$ Gauss and vary the antidot radius
$r_a$. At this value of $B$, only the first two quantum channels in
the probes can propagate.

Fig~(\ref{fig3}) shows the transmission coefficient
$T_{m1},\ (m=1,2,3,4)$ and Hall resistance $R_H$ as a function
of antidot radius $r_a$. Here $T_{m1}$ is the
total transmission coefficient from probe I to probe $m$ including 
all the two propagating channels. From Fig.~(\ref{fig3}a) we can find that
there is a regime with intermediate values of $r_a$: $0.45W<r_a<0.60W$,
where the relations of Eq. (\ref{eq4}) indeed roughly hold.
Thus, just as what happened for the structure of Fig.~(\ref{fig1}), using the
B\"uttiker formula we arrive at the abnormal Hall plateau 
$R_H\approx 1\times h/e^2$ for this range of $r_a$, where both 
propagating channels are only partially transmitting. 
Furthermore, inspecting Fig.~(\ref{fig4}) which plots the transmission
coefficients of individual propagating channels, we see that relations
of Eq. (\ref{eq5}) are also approximately satisfied.  Here 
since $T_{31}^i$ and $T_{41}^i$ are almost zero in the crossover regime, 
they are not shown in Fig.~(\ref{fig4}).

To test the micro-symmetry relations Eq. (\ref{eq7}) for the present
structure, in Table~(1) various transmission probabilities are tabulated for
a given magnetic field $B=6060$ Gauss for several values of the
antidot radius $r_a$.  Clearly our numerical data are consistent with 
Eqs. (\ref{eq7}), which are supported by the general topological equivalence 
argument\cite{lei1}. We recall that the equal signs in Eqs. (\ref{eq7}) are
only rigorously satisfied by the limiting cases of perfectly transmitting or
reflecting transport channels, as shown in our earlier work\cite{lei1}.
In this sense the consistency of our data in Table~(1) with Eq. (\ref{eq7}) 
is quite respectable. Thus we may conclude that in the crossover regime 
of the structure shown in Fig.~(\ref{fig2}), our topological argument and 
the micro-symmetry still approximately hold, and these features
do not depend strongly on the shapes of the Hall junction, at least as far 
as the kind of changes to the Hall junction we have made here, and an 
abnormal Hall plateau can still be observed due to these features. 

On the other hand, there are several features which do depend on the details
of the scattering junction, as shown in Figs.~(\ref{fig3}) and (\ref{fig4}). 
First, for a small antidot {\it e.g.} $0<r_a<0.45W$ and at this field 
$B=6060$ Gauss, we don't have two perfect transmission channels as what was 
found\cite{lei1} for the structure of Fig. (\ref{fig1}). $i.e.$, instead of
$T_{21}=2$ and $T_{11}=0$, we now have $T_{21}\approx 1.05$, 
$T_{11}\approx 0.77$, $T_{31}\approx 0.15$, and $T_{41}\approx 0.03$. 
Second, the transition from the crossover regime to the regime dominated by
AB oscillations (large antidot) is much sharper here.
Finally, the circular-shaped quantum dot gives rise to
relatively more Hall resistance fluctuations than that of the square-shaped
quantum dot studied before\cite{lei1}.  However these features which
depend on the particular shapes of the scattering junction, are
all related to the values of transmission coefficients, which, 
as was well understood in the literature, are expected to be sensitive to 
details of the scattering.

The above numerical results approximately confirms that the appearance 
of crossover regime and the correctness of Eqs.~(\ref{eq7}), (\ref{eq5}) 
and (\ref{eq4}) are quite universal in the sense that they do not depend 
strongly on the particular quantum dot structure we choose. Furthermore, 
our topological explanation of the micro-symmetry in the crossover 
regime~\cite{lei1} does not rely on the exact position where the antidot is 
positioned.  Nevertheless so far the structures studied all had the antidot
positioned at the center of the quantum dot making the systems 4-fold
rotational symmetric. Hence to further establish the transport universality 
of the crossover regime, in the following we investigate situations where 
the antidot is located away from the center.

\subsection{Asymmetric junction}

Consider the Hall junction illustrated in Fig.~(\ref{fig5}a).
All system parameters are the same as before except that the center 
of the antidot is now located at the position ($-W/10,-W/10$) 
(assuming the center of the whole structure is at (0,0)). The
magnetic field is fixed at $B=6060$ Gauss which belongs to 
the crossover regime field strength.

Our numerical calculation seems to provide a positive answer to the question
of universality.  Fig.~(6a1) shows the transmission coefficient
$T_{m1},~(m=1,2,3,4)$ which unambiguously demonstrates that 
Eq.~(\ref{eq4}) still holds here. To test micro-symmetry relations
and topological explanation of Ref. \onlinecite{lei1}, 
Fig.~(6a2) and Table~(2a) show that Eqs.~(\ref{eq5}) and 
(\ref{eq7}) are indeed satisfied.  

Since the center of the antidot is a 4-fold symmetric point inside the
quantum dot, we need to calculate the transmission coefficients of all the
other three cases (Fig.~(\ref{fig5}b,5c,5d)) in order to compute the Hall 
resistance $R_H$ using the B\"{u}ttiker formula Eq.~(\ref{eq1}). Obviously 
the micro-symmetry relations of the crossover regime should hold for
all these situations. Indeed, Figs.~(6b,6c,6d)
and Tables~(2b,2c,2d) confirm that Eqs.~(\ref{eq4}), (\ref{eq5}) and 
(\ref{eq7}) are satisfied for all the cases.  Hence the generic behavior of
the crossover regime has no dependence on the position of the antidot.

Now let us calculate the Hall resistance $R_H$ for the structure of
Fig.~(\ref{fig5}a) when electrons are incoming from probe I. 
Making use of the spatial topological symmetry of the four structures in
Fig.~(\ref{fig5}), the transmission probabilities with electrons 
coming from probes other than probe I of one structure can be obtained 
by a permutation of the transmission probabilities with electrons coming 
from probe I of other structures. In particular, for the transmission 
coefficients we find
\begin{equation}
T_{mn}^{(\alpha)}\ =\ T_{m+5-n,1}^{(\alpha +5-n)}\ ,\ \ (m,n=1,2,3,4;\
\alpha=a,b,c,d)\ \ ,
\label{eq11}
\end{equation}
where $T_{mn}^{(\alpha)}$ is the transmission coefficient from probe $n$ to
probe $m$ of structure $\alpha$ in Fig.~(\ref{fig5}) and where the
index of $T_{mn}^{(\alpha)}$ is taken to be modulo $4$.  For the sake of 
discussion we employ another form of B\"{u}ttiker formula (equivalent 
to Eq.~(\ref{eq1}))~\cite{buttiker3},
\begin{equation}
R_{mn,kl}\ =\ \frac{\alpha_{21}}{\alpha_{11}\alpha_{22}\ -\
\alpha_{12}\alpha_{21}}\ \ ,
\label{eq12}
\end{equation}
where
\begin{equation}
\alpha_{11}\ =\ \frac{e^2}{h}[(N-T_{mm})S\ -\
(T_{ml}+T_{mk})(T_{lm}+T_{km})]/S\ ,
\label{eq13}
\end{equation}
\begin{equation}
\alpha_{12}\ =\ \frac{e^2}{h}[T_{mk}T_{nl}\ -\ T_{ml}T_{nk}]/S\ ,
\label{eq14}
\end{equation}
\begin{equation}
\alpha_{21}\ =\ \frac{e^2}{h}[T_{km}T_{ln}\ -\ T_{lm}T_{kn}]/S\ ,
\label{eq15}
\end{equation}
\begin{equation}
\alpha_{22}\ =\ \frac{e^2}{h}[(N-T_{kk})S\ -\
(T_{km}+T_{kn})(T_{nk}+T_{mk})]/S\ ,
\label{eq16}
\end{equation}
and
\begin{equation}
S\ =\ T_{mk}+T_{ml}+T_{nk}+T_{nl}\ =\ T_{km}+T_{lm}+T_{kn}+T_{ln}\ \ .
\label{eq17}
\end{equation}
Here $N$ is the number of quantum channel occupied by the incident electron,
in our case $N=2$.

Fig.~(\ref{fig7}) shows the dependence of Hall resistance $R_H$ on the
radius $r_a$ of the antidot corresponding to the four Hall-bar structures of 
Fig.~(\ref{fig5}), respectively. It is obvious that the peculiar crossover
regime still exists in every situation,  signaled by the appearance of a
Hall ``plateau'' as $r_a$ is increased into the crossover regime.  We emphasis
that this abnormal plateau is not due to the establishment of perfectly
transmitting edge states when quantum Hall regime is reached.  They are due to
the micro-symmetry properties of the crossover regime.
Hence within the scope of our investigations these properties are universal
against the changes of the antidot position.

Inspecting Fig.~(\ref{fig7}) carefully, we find that the behavior 
of Hall resistance of structure (a) and (c) are quite similar, and so
are the ones of structures (b) and (d). Since structures (a) and (c), 
(b) and (d) are spatial reversal symmetric, it seems that an 
interesting property is obtained,
\begin{equation}
R_H(\mbox {\boldmath{$r$}})\ \approx\ R_H(-\mbox {\boldmath{$r$}})\ ,
\label{eq18}
\end{equation}
where \mbox {\boldmath{$r$}} is the position of the center of the antidot.

In fact, this interesting feature is a natural result of the permutation 
relation Eq.~({\ref{eq11}) and the B\"{u}ttiker formula Eq.~(\ref{eq12}). 
With the help of Eqs.~(\ref{eq11})--(\ref{eq17}), it is rather 
straight forward to obtain the following relations
\begin{equation}
S^{(a)}\ =\ S^{(c)}\ ,\ \ \alpha^{(a)}_{12}\ =\ \alpha^{(c)}_{12}\ ,
\ \ \alpha^{(a)}_{21}\ =\ \alpha^{(c)}_{21}\ \ ,
\label{eq19}
\end{equation}
\begin{equation}
S^{(b)}\ =\ S^{(d)}\ ,\ \ \alpha^{(b)}_{12}\ =\ \alpha^{(d)}_{12}\ ,
\ \ \alpha^{(b)}_{21}\ =\ \alpha^{(d)}_{21}\ \ .
\label{eq20}
\end{equation}
Here the superscripts $(a,b,c,d)$ indicate the structures of 
Fig.~(\ref{fig5}), {\it i.e.} $\alpha_{12}^{(a)}$ is the $\alpha_{12}$ for
structure of Fig.~(\ref{fig5}a). Furthermore, we notice that except two
transition regimes (one is from small-antidot regime to crossover regime and
the other is from crossover regime to large-antidot regime), the behavior of
transmission coefficients of structures (a) and (d) are much alike and so are
those of structures (b) and (c), $i.e,$,
\begin{equation}
T_{m1}^{(a)}\ \approx\ T_{m1}^{(d)}\ ,
\ \ T_{m1}^{(b)}\ \approx\ T_{m1}^{(c)}\ ,\ \ 
(m=1,2,3,4)\ \ .
\label{eq21}
\end{equation}
Consequently we find
\begin{equation}
\alpha_{11}^{(a)}\alpha_{22}^{(a)}\ \approx\ \alpha_{11}^{(c)}
\alpha_{22}^{(c)}\ ,\ \ 
\label{eq22}
\end{equation}
\begin{equation}
\alpha_{11}^{(b)}\alpha_{22}^{(b)}\ \approx\ \alpha_{11}^{(d)}
\alpha_{22}^{(d)}\ .\ \ 
\label{eq23}
\end{equation}
Based on Eqs.~(\ref{eq20})--(\ref{eq23}), it is trivial to obtain
the center-reversal property of $R_H$ (eq.~(\ref{eq18})).

\section{Summary}

In this work we have studied the universality of the micro-symmetry of
transmission coefficients for the peculiar crossover transport regime
established when both quantum Hall and AB effects compete.  This crossover
regime is reached when the antidot radius is increased.
For a square quantum dot with the antidot located in the center, the
crossover regime is marked by the appearance of a host of non-trivial
symmetries between various transmission probabilities.  The consequence
of this set of symmetries is the observation of an abnormal Hall plateau
before the true quantum Hall regime is reached.
We have established the universality of this micro-symmetry from two
directions. First, it is confirmed numerically that for a circular Hall
junction there is also a crossover regime as the antidot size is increased
where the micro-symmetry of transmission probabilities approximately holds. 
Second, when the 4-fold rotational symmetry is broken by shifting the antidot
positions, the same crossover behavior is obtained.  Since, as we have shown
in Ref. \onlinecite{lei1}, that the micro-symmetry relations are supported
by the topological equivalence argument of B\"uttiker\cite{buttiker1}, our
numerical calculations presented here give further confirmation of that
argument.  We may thus conclude that the behavior of the crossover
transport regime, namely the set of micro-symmetry relations, are generic and
robust against the change of shapes of the scattering junction. 

Finally we wish to emphasis that the universality of the crossover behavior 
is established here from a {\it numerical ``experiment''}.  Hence it will
be useful but very challenging to analytically derive the formula for 
transmission coefficients in order to definitely prove the micro-symmetry 
relations.  Furthermore, our numerical calculations were carried out
for system parameters such that only two propagating channels are possible,
which is the simplest case where the topological equivalent argument could
be applied\cite{lei1}.  If there are more propagating channels, the
micro-symmetry relations may become more complicated and more interesting. 

\section{Acknowledgments}

We thank Ningjia Zhu and Haiqing Wei for many useful discussions.
We gratefully acknowledge support from the Natural Sciences and Engineering 
Research Council of Canada,  and le Fonds pour la Formation de Chercheurs 
et l'Aide \`a la Recherche de la Province du Qu\'ebec.

\newpage
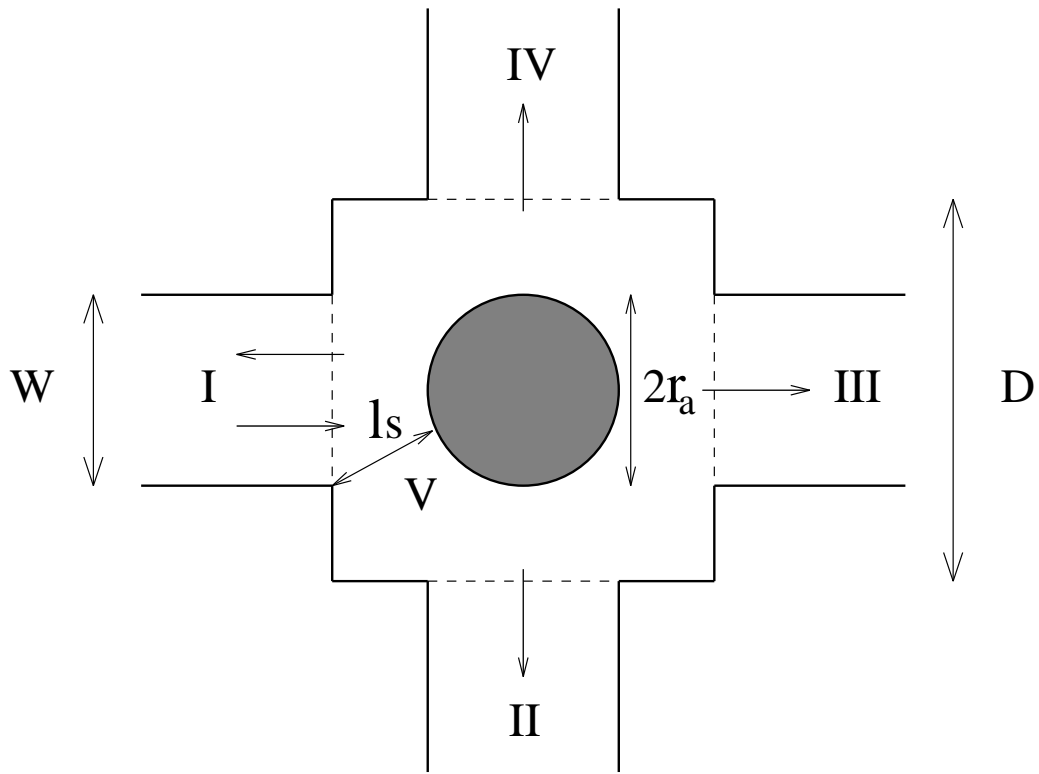
\begin{figure}
\caption{Schematic plot of the Hall junction studied in Ref. [16].
An antidot of radius $r_a$ is confined inside the square quantum dot and
positioned at the center.  Electrons incident from probe I.
}
\label{fig1}
\end{figure}

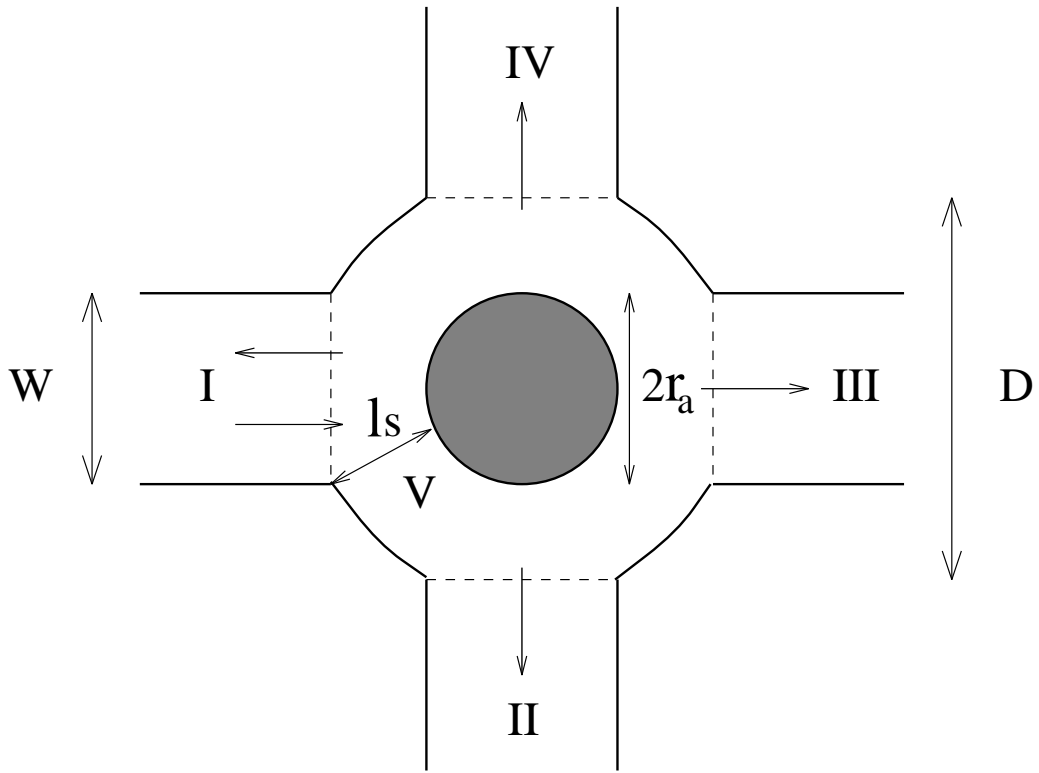
\begin{figure}
\caption{Schematic plot of a circular Hall junction.  An antidot of 
radius $r_a$ is confined inside the junction at the center.  
Electrons incident from probe I.
}
\label{fig2}
\end{figure}

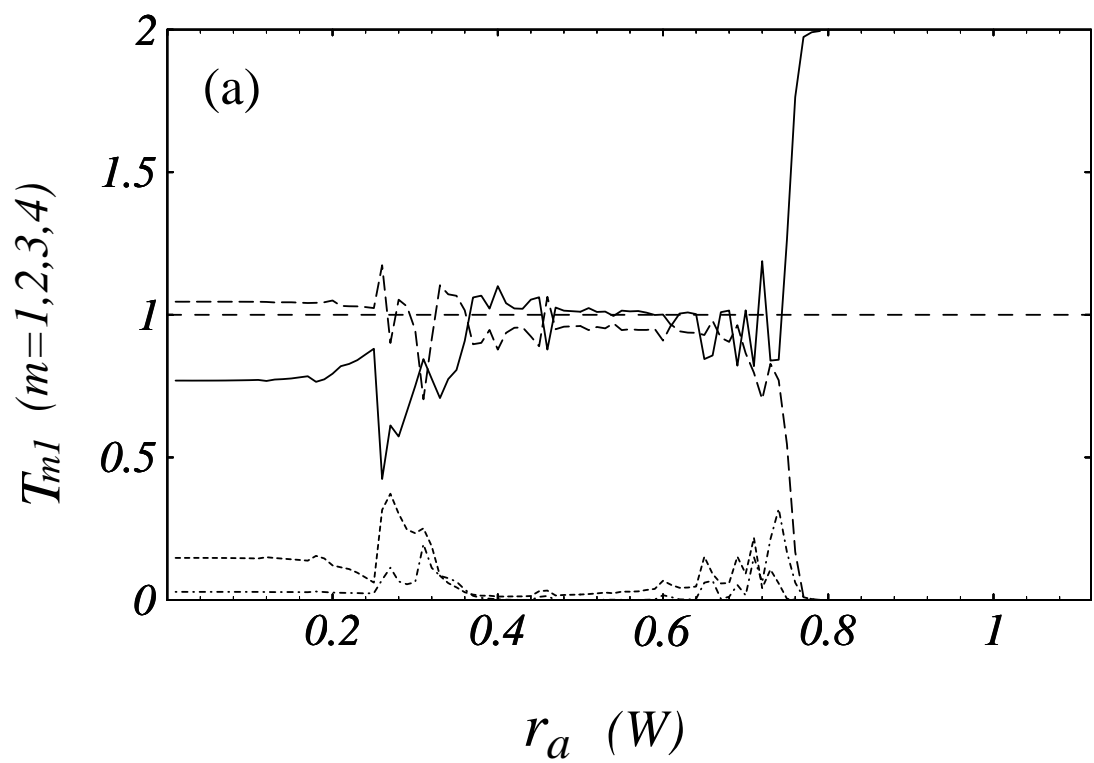
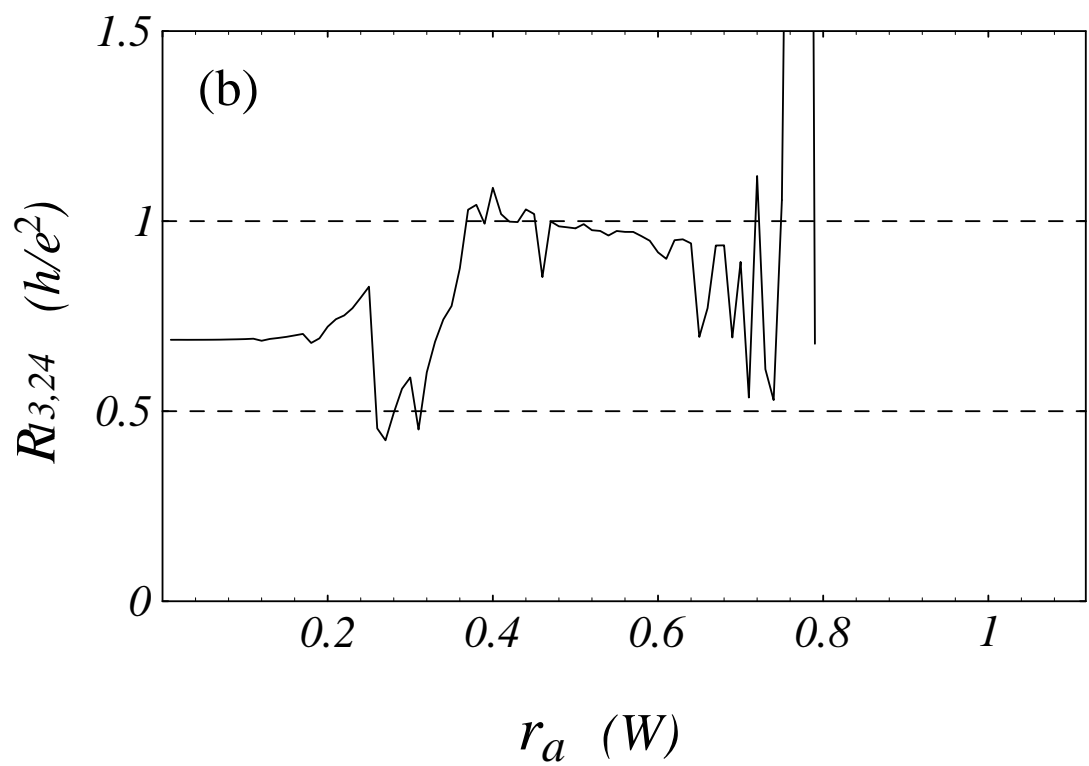
\begin{figure}
\caption{(a) Transmission coefficients $T_{mn}$ of the circular junction
in Fig. (\ref{fig2}) as a function of the antidot radius $r_a$ at 
$B=6060$ Gauss. Note in the crossover regime $T_{11}\approx T_{21} \approx 1$
while $T_{31}\approx T_{41}\approx 0$. Solid line is $T_{11}$, dashed line
is $T_{21}$, short-dashed line is $T_{31}$ and dot-dashed line is $T_{41}$.
(b) Hall resistance $R_H$ as a function of $r_a$ at $B=6060$
Gauss. The crossover regime is marked by the abnormal Hall ``plateau'' 
$\sim h/e^2$.
}
\label{fig3}
\end{figure}

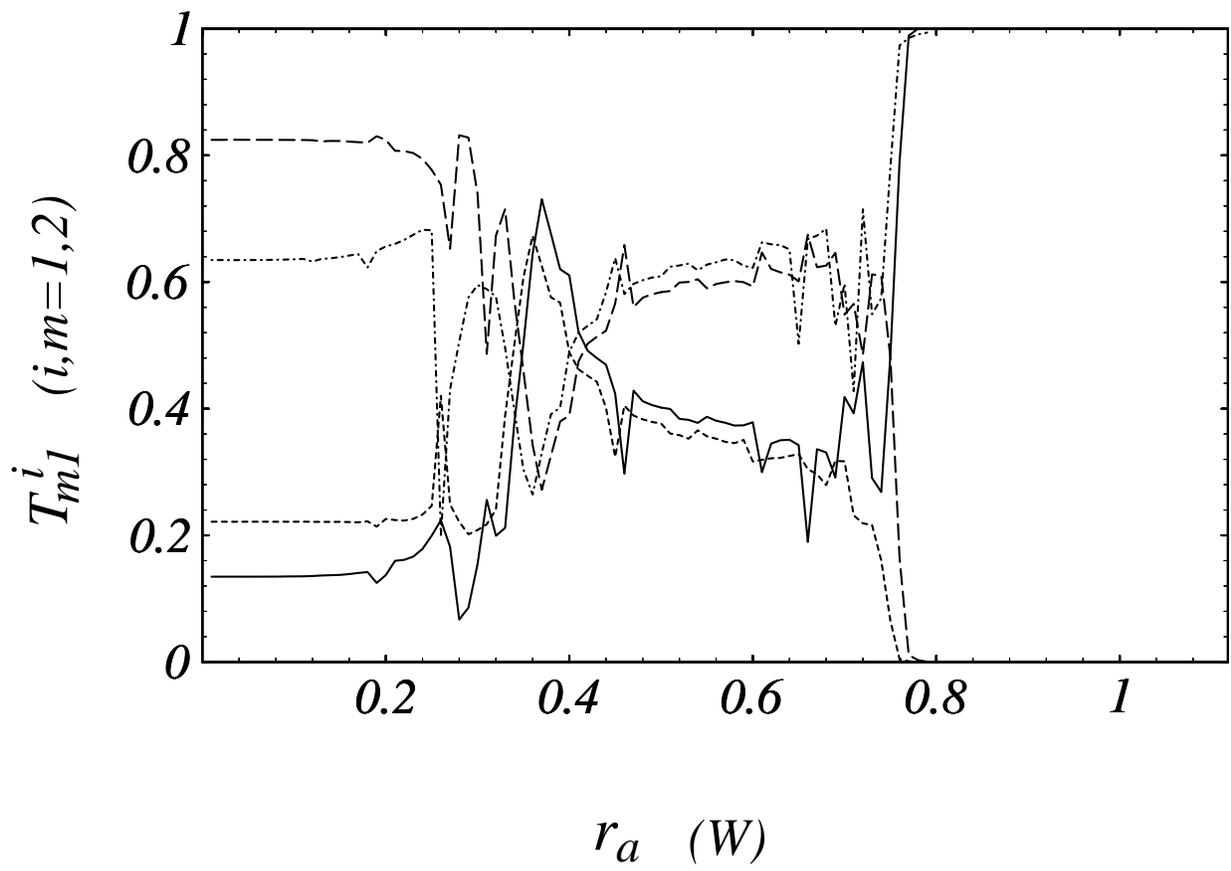
\begin{figure}
\caption{Transmission coefficients of the circular Hall junction for 
individual incoming channels $T_{mn}^i$, where $i$ is the channel number, 
as a function of $r_a$ at $B=6060$ Gauss. Note in the crossover regime 
$T_{11}^1\approx T_{21}^2$, $T_{21}^1\approx T_{11}^2$. Solid line is 
$T_{11}^1$, dashed line is $T_{21}^1$, short-dashed line is $T_{21}^2$ 
and dot-dashed line is $T_{11}^2$.
}
\label{fig4}
\end{figure}

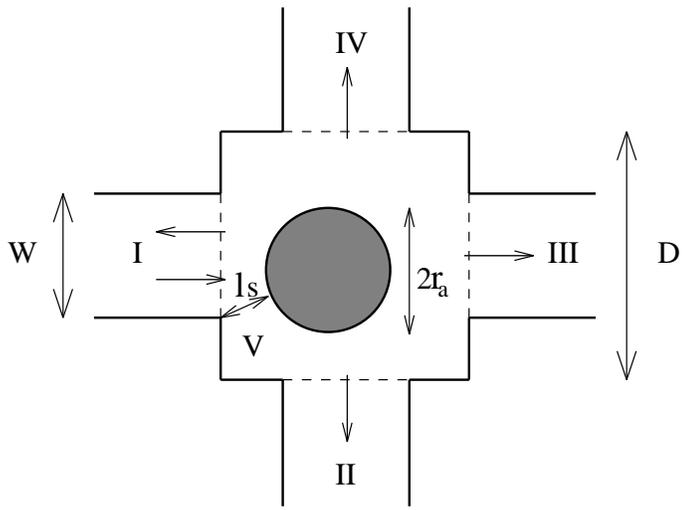
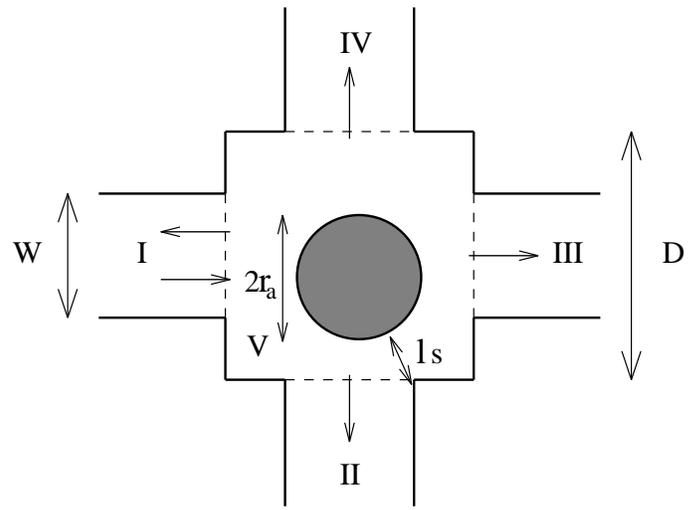
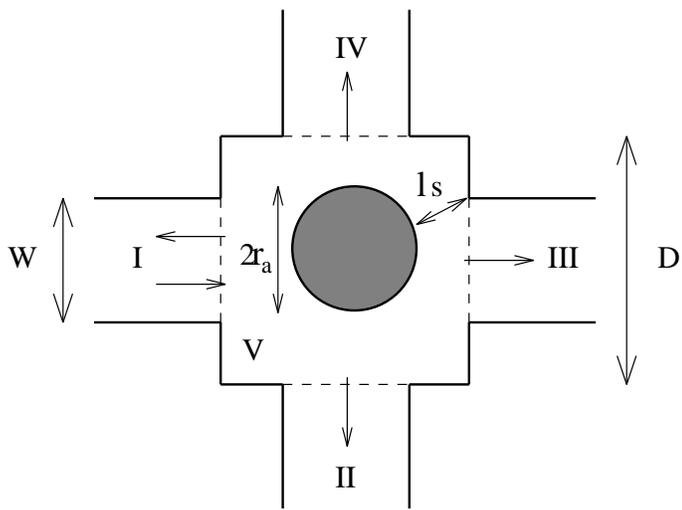
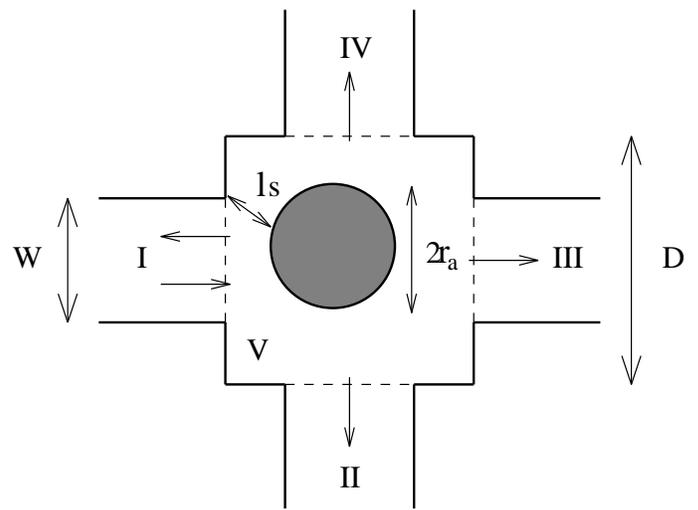
\begin{figure}
\caption{Schematic plot of the Hall junctions with the antidot away from the
center.  Electrons incident from probe I. The centers of the antidots of 
structure (a), (b), (c) and (d) are located at positions ($-W/2,-W/2$), 
($W/2,-W/2$), ($W/2,W/2$) and ($-W/2,W/2$) respectively.
$W$ is the width of the probes and point (0,0) is the center of the 
Hall junctions.
}
\label{fig5}
\end{figure}

\begin{figure}
\caption{Transmission coefficients of the structure shown in 
Fig.~(5a,5b,5c,5d). (a1,b1,c1,d1): Transmission coefficients $T_{mn}$ as 
a function of $r_a$ at $B=6060$ Gauss.  Note that in the crossover regime 
$T_{11}=T_{21}=1$ while $T_{31}=T_{41}=0$. Solid line is $T_{11}$, 
dashed line is $T_{21}$, short-dashed line is $T_{31}$ and dot-dashed line is
$T_{41}$.  (a2,b2,c2,d2): Transmission coefficients of individual
incoming channels $T_{mn}^i$ as a function of $r_a$ at $B=6060$ Gauss. 
Note in the crossover regime $T_{11}^1=T_{21}^2$, $T_{21}^1=T_{11}^2$. 
Solid line is $T_{11}^1$, dashed line is $T_{21}^1$, short-dashed line 
is $T_{21}^2$ and dot-dashed line is $T_{11}^2$.
}
\label{fig6}
\end{figure}

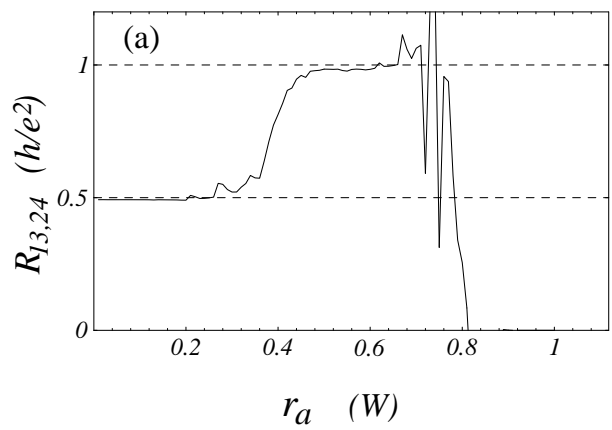
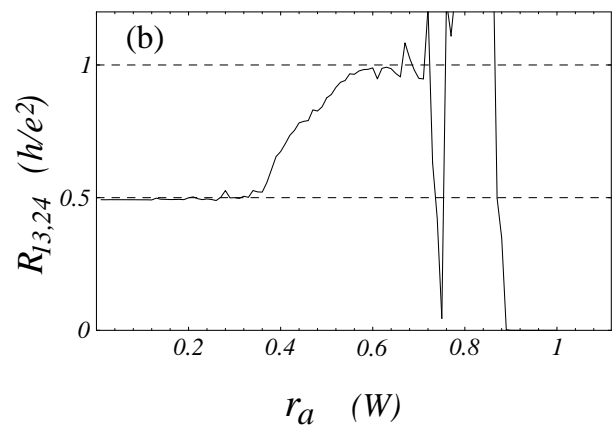
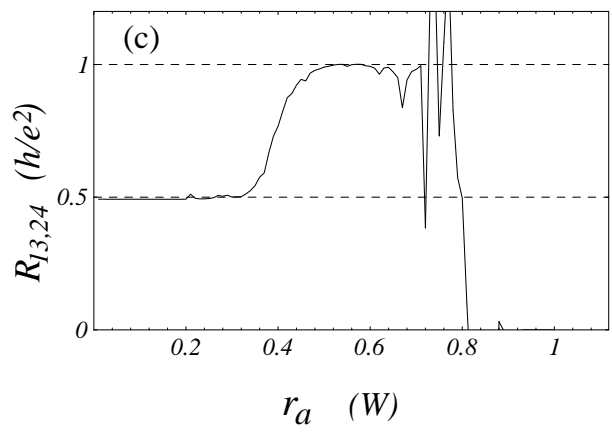
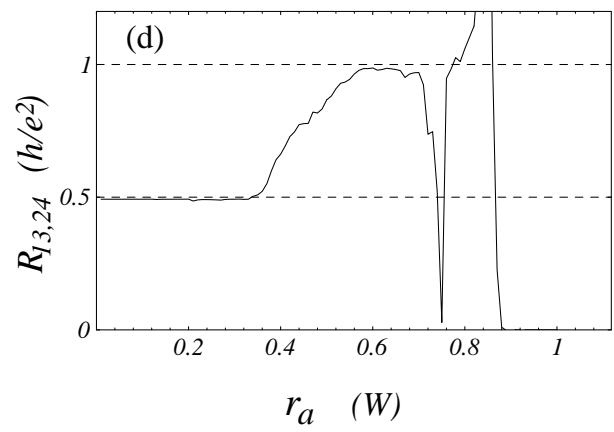
\begin{figure}
\caption{Hall resistance $R_H$ for the four structures in Fig. (\ref{fig5})
as a function of $r_a$ at $B=6060$ Gauss. The crossover regime is marked 
by the abnormal Hall ``plateau'' of $\sim h/e^2$. (a), (b), (c) and (d) are 
for the junctions Fig.~(\ref{fig5}a, 5b, 5c and 5d) respectively.
}
\label{fig7}
\end{figure}

\newpage
\noindent
Table~(1). Transmission coefficients $T_{ij,mn}$ of the circular-shaped Hall
junction for different antidot size $r_a$ at $B=6060$ Gauss. The data 
shows approximately the micro-symmetry of Eq.~(\ref{eq7}).
\vspace{1.0in}

\begin{center}
\begin{tabular}{|c|cccc|}
\multicolumn{5}{c}{}\\ \hline\hline
&         &$B=6060 Gauss$  & &\\
$T_{ij,mn}$  &  $r_a$=0.45$W$  &  $r_a$=0.50$W$  &  $r_a$=0.55$W$  &
$r_a$=0.60$W$ \\ \hline
$T_{11,11}$    & 0.19625 & 0.16980 & 0.16338 & 0.15926  \\
$T_{22,21}$    & 0.15906 & 0.14777 & 0.13730 & 0.14367  \\ \hline
$T_{22,11}$    & 0.35758 & 0.38569 & 0.39566 & 0.39751  \\
$T_{11,21}$    & 0.36527 & 0.37547 & 0.37030 & 0.36799  \\ \hline
$T_{21,11}$    & 0.22735 & 0.22241 & 0.22440 & 0.23028  \\
$T_{12,21}$    & 0.25799 & 0.24585 & 0.24189 & 0.22988  \\ \hline
$T_{12,11}$    & 0.22701 & 0.22224 & 0.22432 & 0.22024  \\
$T_{21,21}$    & 0.19286 & 0.21024 & 0.23071 & 0.24458  \\
\hline
\end{tabular}
\end{center}

\newpage

\noindent
Table~(2). Transmission coefficients $T_{ij,mn}$ of square-shaped Hall
junction for different antidot size $r_a$ at $B=6060$ Gauss. The data 
shows approximately the micro-symmetry of Eq.~(\ref{eq7}). The four tables
labeled (a)--(d) correspond to the four cases of Fig.~(5).
\vspace{0.5in}

\begin{center}
\begin{tabular}{|c|cccc|}
\multicolumn{5}{c}{}\\ \hline\hline
(a) &   &$B=6060 Gauss$  & &\\
$T_{ij,mn}$  &  $r_a$=0.45$W$  &  $r_a$=0.50$W$  &  $r_a$=0.55$W$  &
$r_a$=0.60$W$ \\ \hline
$T_{11,11}$    & 0.05562 & 0.04323 & 0.05066 & 0.02705  \\
$T_{22,21}$    & 0.06449 & 0.05176 & 0.11119 & 0.02531  \\ \hline
$T_{22,11}$    & 0.51773 & 0.57690 & 0.50621 & 0.74003 \\
$T_{11,21}$    & 0.50523 & 0.58387 & 0.51868 & 0.70998  \\ \hline
$T_{21,11}$    & 0.22367 & 0.20044 & 0.27021 & 0.14702  \\
$T_{12,21}$    & 0.20345 & 0.18064 & 0.17242 & 0.13588  \\ \hline
$T_{12,11}$    & 0.18651 & 0.17520 & 0.17257 & 0.08857  \\
$T_{21,21}$    & 0.22118 & 0.17618 & 0.15352 & 0.11026  \\
\hline
\end{tabular}
\end{center}

\begin{center}
\begin{tabular}{|c|cccc|}
\multicolumn{5}{c}{}\\ \hline\hline
(b) &   &$B=6060 Gauss$  & &\\
$T_{ij,mn}$  &  $r_a$=0.50$W$  &  $r_a$=0.55$W$  &  $r_a$=0.60$W$  &
$r_a$=0.65$W$ \\ \hline
$T_{11,11}$    & 0.07671 & 0.08561 & 0.05593 & 0.03802  \\
$T_{22,21}$    & 0.06021 & 0.04987 & 0.02631 & 0.01327  \\ \hline
$T_{22,11}$    & 0.42673 & 0.46831 & 0.54855 & 0.61120 \\
$T_{11,21}$    & 0.50997 & 0.50551 & 0.59186 & 0.64374  \\ \hline
$T_{21,11}$    & 0.21832 & 0.27954 & 0.22030 & 0.17901  \\
$T_{12,21}$    & 0.22536 & 0.26623 & 0.23354 & 0.18041  \\ \hline
$T_{12,11}$    & 0.17649 & 0.16623 & 0.17363 & 0.16191  \\
$T_{21,21}$    & 0.16720 & 0.13305 & 0.13545 & 0.11159  \\
\hline
\end{tabular}
\end{center}

\newpage

\begin{center}
\begin{tabular}{|c|cccc|}
\multicolumn{5}{c}{}\\ \hline\hline
(c) &   &$B=6060 Gauss$  & &\\
$T_{ij,mn}$  &  $r_a$=0.60$W$  &  $r_a$=0.625$W$  &  $r_a$=0.65$W$  &
$r_a$=0.675$W$ \\ \hline
$T_{11,11}$    & 0.05560 & 0.07254 & 0.03809 & 0.04321  \\
$T_{22,21}$    & 0.07046 & 0.09014 & 0.04203 & 0.03209  \\ \hline
$T_{22,11}$    & 0.51735 & 0.46176 & 0.61005 & 0.61994  \\
$T_{11,21}$    & 0.51742 & 0.45744 & 0.61337 & 0.65171  \\ \hline
$T_{21,11}$    & 0.21657 & 0.22544 & 0.16602 & 0.14622  \\
$T_{12,21}$    & 0.18755 & 0.19846 & 0.15711 & 0.15401  \\ \hline
$T_{12,11}$    & 0.20810 & 0.23056 & 0.17426 & 0.18120  \\
$T_{21,21}$    & 0.21923 & 0.25808 & 0.18054 & 0.14963  \\
\hline
\end{tabular}
\end{center}

\vspace{0.5in}

\begin{center}
\begin{tabular}{|c|cccc|}
\multicolumn{5}{c}{}\\ \hline\hline
(d) &   &$B=6060 Gauss$  & &\\
$T_{ij,mn}$  &  $r_a$=0.45$W$  &  $r_a$=0.50$W$  &  $r_a$=0.55$W$  &
$r_a$=0.60$W$ \\ \hline
$T_{11,11}$    & 0.05564 & 0.04329 & 0.05162 & 0.02766  \\
$T_{22,21}$    & 0.06345 & 0.06690 & 0.07459 & 0.02671  \\ \hline
$T_{22,11}$    & 0.51416 & 0.51615 & 0.48591 & 0.73881  \\
$T_{11,21}$    & 0.49947 & 0.48176 & 0.46726 & 0.68993  \\ \hline
$T_{21,11}$    & 0.20006 & 0.18348 & 0.17073 & 0.08608  \\
$T_{12,21}$    & 0.19157 & 0.15113 & 0.14174 & 0.07760  \\ \hline
$T_{12,11}$    & 0.21018 & 0.25320 & 0.28065 & 0.19681  \\
$T_{21,21}$    & 0.24069 & 0.29544 & 0.30648 & 0.24394  \\
\hline
\end{tabular}
\end{center}

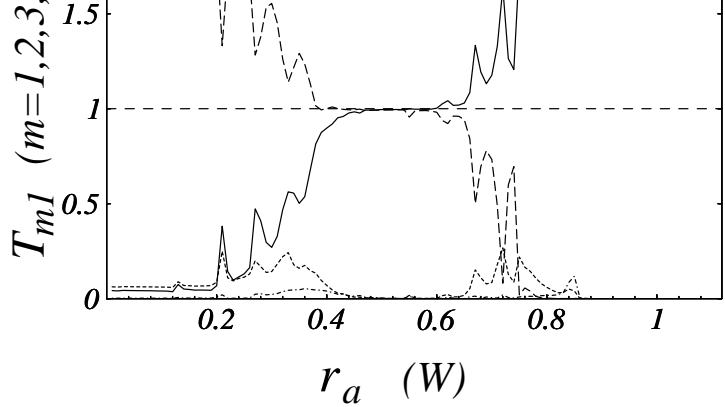
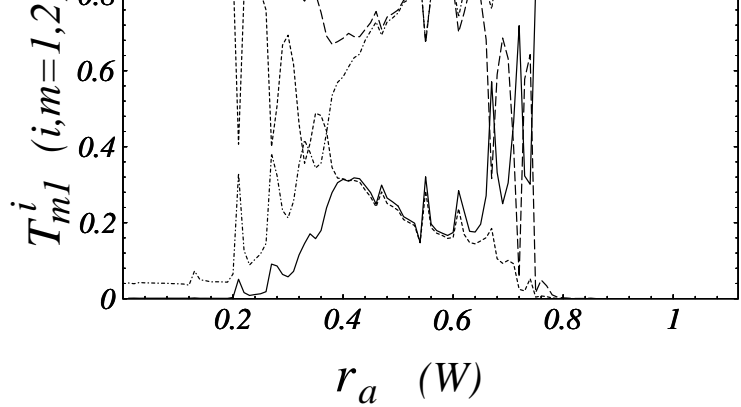
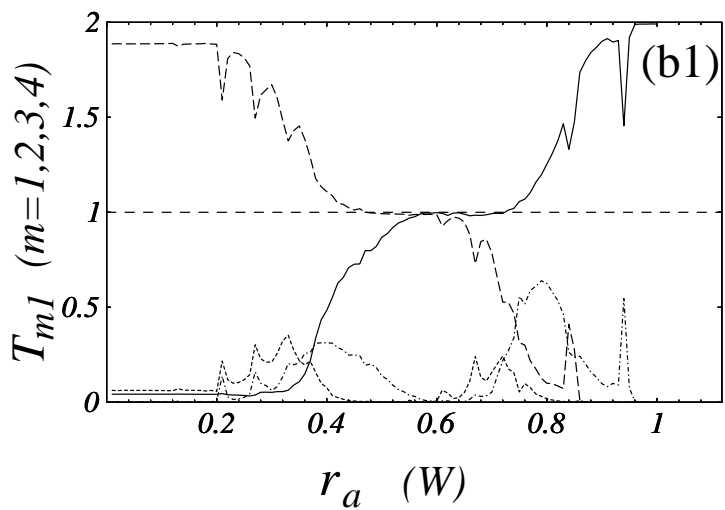
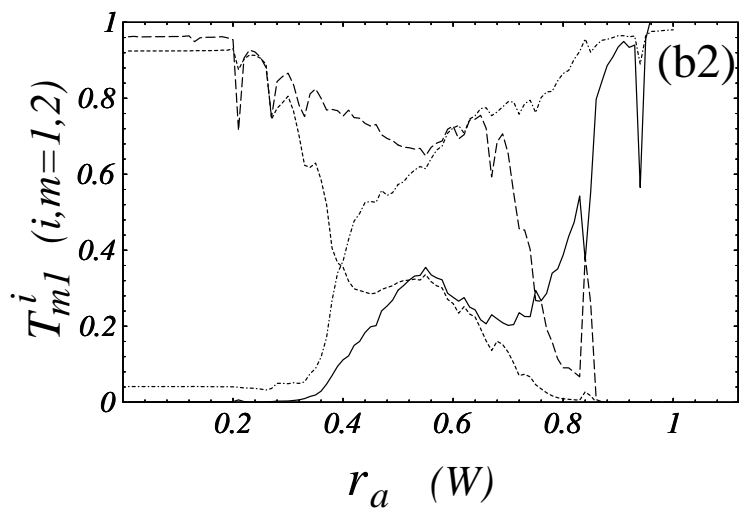
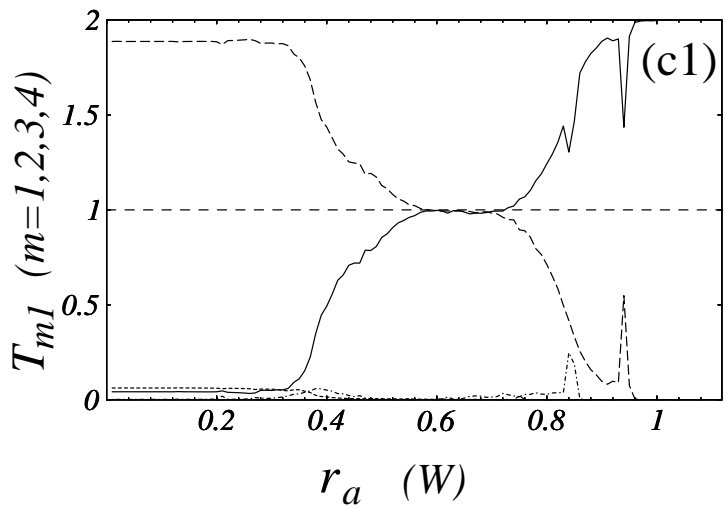
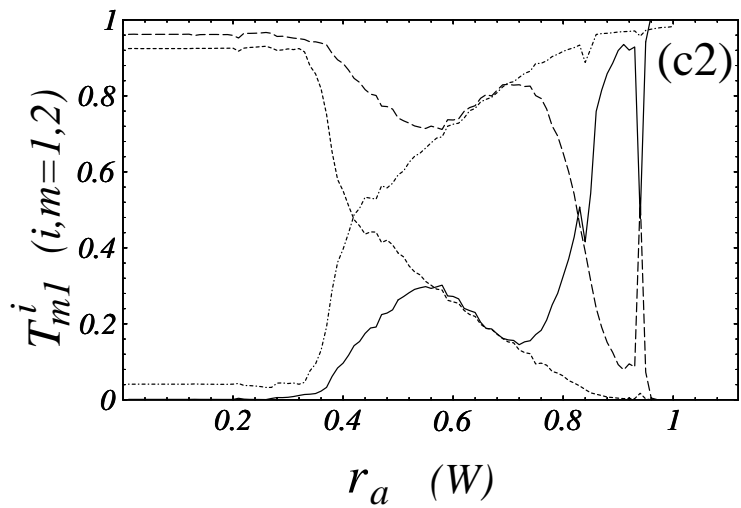
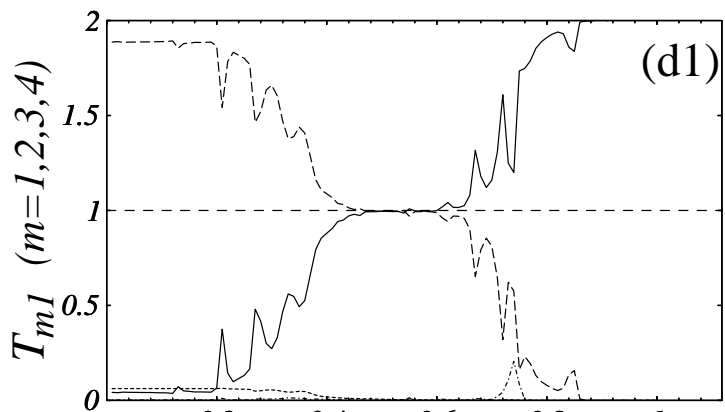
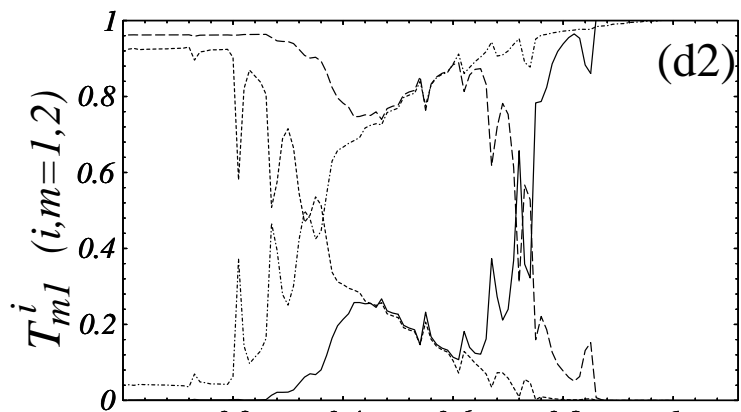
\end{document}